\documentstyle[12pt]{article}
 
 \title{ Fractional Spin through Quantum Affine Algebra   $\hat A(n) $ and quantum affine superalgebra  $\hat A(n,m) $ } 
 \author{\bf { M. Mansour$^{1}$ \thanks{email: mansour70@mailcity.com}
  and ~ M. Daoud$^{1,2}$} \\ 
 $^{1}$ Laboratoire de Physique Th\'eorique, Facult\'e des Sciences\\~B. P. 1014 Rabat.  Morocco.\\ 
 $^{2}$ D\'epartement de Physique, 
 Facult\'e des Sciences. B. P. 28/S \\ ~Universit\'e Ibnou Zohr.  Agadir.   Morocco.}
 
\begin{document}           
 \maketitle              
 $$\bf{Abstract}$$
 Using the splitting of a $Q$-deformed boson, in 
 the $Q \to q= e^{\frac{\rm 2\pi i}{\rm  k}}$ limit, 
 the fractional decomposition of the quantum affine  algebra
  $\hat A(n) $ and the quantum affine superalgebra $\hat A(n,m) $ are found.  This
  decomposition is based on the oscillator representation
  and can be related to the fractional supersymmetry and 
 k-fermionic spin. We establish also 
 the equivalence between the  quantum affine algebra   $\hat A(n) $
  and the classical one in the fermionic 
 realization. 
 
 \vfill
 \pagebreak
 \section{Introduction}
  The concept of quantum group and algebra \cite{r1,r2}, has extensively entered mathematical and physical literatures. In theoretical physics, quantum groups have received considerable interest due to their connection with quantum Yang-Baxter equation \cite{r23} and the quantum inverse scattering method \cite{r24}. From a mathematical point of view, quantum groups and algebras can be regarded as deformations of  the universal enveloping algebras of  semi-simple Lie algebras. The quantum analogues of  Lie superalgebras has been constructed in \cite{r3,r4}. Quantized enveloping algebras associated to affine algebras and superalgebras are given in \cite{r1,y}. Many properties of quantum groups and algebras are quit similar to or richer than ones of the usual Lie groups and algebras in connection with  the representations theory.
 It is well known that boson realization method is a very powerful and elegant method for the study of quantum algebras representations. Based on this method, the representation theory of quantum affine algebras has been an object of intensive studies; Available are the results for the oscillator representations of affine algebra. There are obtained \cite{f1,f2,f3} through consistent realizations involving deformed Bose and Fermi operators \cite{r9,r10}.\\
 \hspace*{0.5cm}  Recently, in connection with quantum group theory, 
 a new geometric interpretation of fractional supersymmetry  
  has been developed in  \cite{r11,r12,r13,r14,r15}. In these works, the authors 
  show that the one-dimensional superspace is isomorphic to the braided line
  when the deformation parameter goes to a root of unity. Similar techniques are used, in the reference \cite{r18}, to 
  show how internal spin arises naturally in certain limit of the $Q$-deformed
  angular momentum algebra $U_{Q}(sl(2))$. Indeed, using  $Q$-Schwinger 
  realization, it is shown that the decomposition of the
    $U_{Q}(sl(2))$  into a direct product of not deformed $U(sl(2))$ and 
   $U_{q}(sl(2))$ which is the same version of $U_{Q}(sl(2))$ at $Q =q$. The property 
  of splitting of quantum algebras $A_{n},  B_{n},  C_{n} $ and $D_{n}$ and quantum superalgebra $A(m,n),  B(m,n),  C(n+1) $ and $D(n,m)$ in the 
  $ Q \to q$ limit is investigated in \cite{M1}. The case of deformed Virasoro algebra and some other particular quantum (Super)-algebra is given in \cite{r19}.
 
  \hspace*{0.5cm}  The aim of this  paper is to investigate the property 
  of decomposition  of the quantum affine algebra $\hat A(n)$ and the quantum affine superalgebra $\hat A(n,m) $ in the $ Q \to q$ limit. As a first step we wish to present in the next section (section 2) a 
  some  results  concerning k-fermions. In section 3, we discuss the property of $Q$-boson decomposition
  in the $ Q \to q $ limit. We introduce the way in which
  one obtains two independent objects ( an ordinary boson and a $k$-fermion )
   from one $Q$-deformed boson when $Q$ goes to a root of unity. We establish  
   also the equivalence between a $Q$-deformed fermion and a conventional
   (ordinary ) one. Using these results, we analyse the $ Q \to q $  limit of the quantum affine algebra 
 $U_{Q}(\hat A(n))$ (section 4) and the quantum affine superalgebra $\hat A(m,n)$ (section 5). We note that Q-oscillator realization is crucial in this paper. Therefore, the results obtained in this work are valid for the oscillator representations. In a last section (section 6) we shall give some concluding remarks .
   \section{Introducing $k$-fermionic algebra.}
 The usual starting commutation relations of $q$-deformed bosonic algebra $\sum_q$ are
 
 \begin{equation} \label{e1}
 \begin{array}{c}
 a^- a^+ -qa^+ a^-~=~q^{-N}\\
 a^{-} a^{+} -q^{-1}a^{+} a^{-}~=~q^{N}\\
 q^Na^+q^{-N}~=~qa^+\\
 q^Na^-q^{-N}~=~q^{-1}a^-\\
 q^Nq^{-N}~=~q^{-N}q^N~=1~
 \end{array}
 \end{equation}
 where the deformation parameter
 \begin{equation} \label{e2}
 q~=~exp(\frac{2 \pi i}{l})~~~~~~~ l \in N -\{0,1\} 
 \end{equation}
 is a root of unity. The annihilation operator $a^-$ is hermitic conjugated to creation operator $a^+$. the number operator $N$ is hermitic. From equation (\ref{e1}), it is not difficult to obtain the following relations
 \begin{equation} \label{e3}
 \begin{array}{c}
 a^-(a^+)^n~=~[{[n]}]q^{-N}(a^+)^{n-1}+q^N(a^+)^na^-\\
 (a^-)^n a^+~=~[{[n]}](a^+)^{n-1}q^{-N}+q^Na^+(a^-)^n\\
 \end{array}
 \end{equation}
 
 Where the symbol $[{[]}]$ is defined by:
 \begin{equation} \label{e4}
 [{[n]}]=\frac{1-q^{2n}}{1-q^2}
 \end{equation}
 The cases of odd and even values of $\ell$ have to be treated in slightly different ways. Hence, we introduce a new variable $k$ defined by
 \begin{equation} \label{e5}
 \begin{array}{c}
 k=l \qquad\hbox{for odd values of} ~~~l\\
 k={l\over 2} \qquad\hbox{for even values of} ~~~l\\
 \end{array}
 \end{equation}
 such that for odd $l$ (rep. even $l$), we have $q^k=1$ (resp.,$ q^k=-1$).
 In the particular case $n=k$, equations $(\ref{e3})$ are  amenable to the form
 \begin{equation} \label{e6}
 \begin{array}{c}
 a^-(a^+)^k~=~\pm(a^+)^k a^-\\
 (a^-)^k a^+~=~\pm a^+(a^-)^k
 \end{array}
 \end{equation}
  In addition, the equation (\ref{e1}) yield
 \begin{equation} \label{e7}
 \begin{array}{c}
 q^N(a^+)^k~=~(a^+)^k q^N\\
 q^N(a^-)^k~=~(a^-)^k q^N
 \end{array}
 \end{equation}
 
 We point out that the elements $(a^+)^k$ and $(a^-)^k$ are elements of the centre of $\sum_q$ algebra (odd~~ $l$). The irreducible representations are $k$-dimensional. Due to the fact that the elements $(a^+)^k$ and $(a^-)^k$ are central, if one deals with a $k$-dimensional representation, we have 
 \begin{equation} \label{e8}
 (a^+)^k~=~\alpha ~~I ,~~~~(a^-)^k~=~\beta ~~I
 \end{equation}
 
  The extra possibilities parameterised by 
 
  $$(i) ~~~~ \alpha = 0 ~~~~\beta \ne 0$$
  $$(ii) ~~~~ \alpha \ne 0~~~~\beta =  0$$
  $$(iii) ~~~~ \alpha \ne 0~~~~\beta \ne 0$$
 
  are not relevant for the considerations of this paper. The case (iii) correspond to the periodic representation and in the cases (i) and (ii) we have the so-called semiperiodic(semicyclic) representation. In what follows, we shall deal with a representation of the algebra $\sum_q$ such that 
 \begin{equation}
 (a^+)^k~=0~,~~~~~~~~~~~~~~(a^-)^k~=0~
 \end{equation}
 are satisfied. We note that the algebra $\sum_{-1}$ obtained for $k=2$, correspond to ordinary fermion operators with $(a^+)^2~=0$ and $(a^-)^2~=0$ which reflects the Pauli exclusion principle. In the limit case where $k\to\infty$, we have the algebra  $\sum_{1}$, which correspond to the ordinary boson operators. For $k$ arbitrary, the algebra  $\sum_{q}$ correspond to the $k-$fermions(or anyons with fractional spin in the sense of Majid \cite{r16,r17}) operators that interpolate between fermion and boson operators.
 \section{Fractional spin through Q-boson.}
 In the previous section, we have been working with $q$ a root of unity. When $q^l=1$, quantum oscillator($k-$fermionic) algebra exhibit rich representation behaviour with very special properties different from the generic case. In the first case the Hilbert space is finite dimensional, wile in the generic case the Fock space is infinite dimensional Hilbert space. Now, let us consider, in order to investigate the decomposition of a $Q-$deformed boson $(q \in C)$  in the $Q \to exp({2\pi~ i\over k})$, the $Q-$deformed algebra $\Delta_Q$. The algebra  $\Delta_Q$ is generated by an annihilation operator $B^-$, a creation operator $B^+$ and a number operator $N_B$ with the relations
 \begin{equation} \label{e9}
 \begin{array}{c}
 B^-B^+-QB^+B^-~=~Q^{-N_B}\\
 B^-B^+-Q^{-1}B^+B^-~=~Q^{N_B}\\
 Q^{N_B}B^+Q^{-N_B}~=~QB^+\\
 Q^{N_B}B^-Q^{-N_B}~=~Q^{-1}B^-\\
 Q^{N_B}Q^{-N_B}~=~Q^{-N_B}Q^{N_B}~=~1
 \end{array}
 \end{equation}
 From equation (\ref{e9}), we obtain
 \begin{equation} \label{e10}
 [Q^{-N_B}B^-,[Q^{-N_B}B^-,[\dots[Q^{-N_B}B^-,(B^+)^k]_{Q^{2k}}\dots]_{Q^{4}}]_{Q^{2}}]=Q^{{k(k-1)}\over{2}}[k]!
 \end{equation}
 Where the $Q-$deformed factorial is given by 
 $$[k]!~=~[k][k-1][k-2]\dots[1]$$
 $$[0]!~=~1$$
 with $$[k]~=~\frac{Q^k-Q^{-k}}{Q^{1}-Q^{-1}}$$
 
 The $Q-$commutator, in (\ref{e10}), of two operators $A$ and $B$ is defined by
 $$[A,B]_Q~=~AB- Q BA$$
  The aim of this section is to determine the limit of the $\Delta_Q$ algebra when $Q$ goes to the root of unity $q$ see
 (\ref{e2}). The starting point is the limit $Q\to q$ of equation (\ref{e10})
 \begin{equation} \label{e11}
 \begin{array}{c}
 lim_{Q\to q} {1\over k}[Q^{-N_B}B^-,[Q^{-N_B}B^-,[\dots[Q^{-N_B}B^-,{B^+}^k]_{Q^{2k}}\dots]_{Q^{4}}]_{Q^{2}}]\\
 =lim_{Q\to q}{Q^{{k(k-1)}\over {2}}\over{[k]!}}[Q^{-N_B}{(B^-)^k}, (B^+)^k]\\
 =q^{k(k-1)\over 2 }.
 \end{array}
 \end{equation}
 
 The equation (\ref{e11}) can be reduced to :
 \begin{equation} \label{e15}
 lim_{Q\to q}[{{{Q^{kN_B\over 2}{(B^-)}^k}\over{([k]!)^{1\over 2}}},{{{(B^+)}^kQ^{{kN_B\over 2}}\over{([k]!)^{1\over 2}}}}}]~=~1
 \end{equation}
 
 We note that since $q$ is a root of unity, it is possible to change the sign on the exponent of $q^{kN_B\over 2}$ terms in the above and in the following definitions (when $Q\to q$).
 
 Following the work[18], we define the operators
 \begin{equation} \label{e16}
 \begin{array}{c}
 b^-~=~lim_{Q\to q}{{Q^{\pm k N_B \over 2}{(B^-)}^k}\over{([k]!)^{1\over 2}}}\\
 b^+~=~lim_{Q\to q}{{{(B^+)}^kQ^{{\pm kN_B\over 2}}\over{([k]!)^{1\over 2}}}}
 \end{array}
 \end{equation}
 
 then we obtain
 \begin{equation} \label{e161}
 [b^-, b^+]~=~1
 \end{equation}
 
 Which are nothing but the commutation relation of an ordinary boson. The number operator of this new bosonic oscillator is defined, in the usual ways, as $ N_b~=~b^+b^-$.\\
 This type of reasoning, concerning the $Q\\to q$ limit of $Q-$boson, has been invoked for the first time in the references [13-16,18] in order to investigate the fractional supersymmetry. (In these references, the authors show that there is an isomorphism between the braided line and one dimensional superspace.)\\
 Now, we are in a position to discuss the splitting of $Q-$deformed boson in the $Q\to q$ limit. Let us  introduce the new set of generators given by:
 
 \begin{equation} \label{e17}
 \begin{array}{c}
 A^-~=~B^-q^{-kN_b\over 2}\\
 A^+~=~B^+q^{-kN_b\over 2}\\
 N_A~=~N_B-k N_b
 \end{array}
 \end{equation}
 
 which satisfie the following commutation relations
 \begin{equation} \label{e21}
 \begin{array}{c}
 \lbrack A^{-}, A^{+} \rbrack_{q^{-1}} =  q^{N_{A}}\\
 \lbrack A^{-}, A^{+} \rbrack_{q}  =  q^{- N_{A}}\\
 \lbrack N_{A}, A^{\pm} \rbrack =\pm A^{\pm}\\
 \end{array}
 \end{equation}
 
 and then define a $k-$fermion. The two algebras generated by the set of operators $\{ b^+,b^-,N_b\}$ and $\{ A^+,A^-,N_A\}$ are mutually commutative. We thus conclude that in the $Q\\to q$ limit, the $Q-$deformed bosonic algebra oscillator decomposes into two independents oscillators, an ordinary boson and $k-$fermion.\\
 There is also a natural question which emerges: is it possible to find similar splitting property for $Q-$deformed fermionic operators when the deformation parameter $Q$ reduce to a root of unity $q$ ? To answer to this question, we consider the $Q-$deformed fermionic algebra generated by the operators $F^-,F^+$ and $N_F$ satisfying the following relations
 
 \begin{equation} \label{e21}
 \begin{array}{c}
 F^-F^+~+~QF^+F^-~=~Q^N\\
 F^-F^+~+~Q^{-1}F^+F^-~=~Q^{-N}\\
 Q^{N_F}F^+Q^{-N_F}~=~QF^+\\
 Q^{N_F}F^-Q^{-N_F}~=~Q^{-1}F^-\\
 Q^{N_F}Q^{-N_F}~=~Q^{-N_F}Q^{N_F}~=1~\\
 (F^+)^2=0,(F^-)^2=0
 \end{array}
 \end{equation}
 We define the new operators
 \begin{equation} \label{e23}
 \begin{array}{c}
 f^-~=~Q^{-N_F\over 2}F^-\\
 f^+~=~F^+Q^{-N_F\over 2}
 \end{array}
 \end{equation}
 We obtain by a direct calculation the following anti-commutation relation
 \begin{equation} \label{e25}
 \{ f^-,f^+\}~=~1
 \end{equation}
 Moreover, we have the nilpotency conditions
 \begin{equation} \label{e26}
 \begin{array}{c}
 (f^-)^2~=~0\\
  (f^+)^2~=~0
 \end{array}
 \end{equation}
 Thus, we see that the $Q-$deformed fermion reproduces the conventional(ordinary)fermion.

 \section{Quantum  affine algebra $ U_{Q}(\hat A(n))$ at $Q$ a root of unity}
 \hspace*{0.5cm} 
  We use now the above results to derive the property of decomposition
 of quantum affine algebras $ U_{Q}(\hat A(n))$ in the $Q \to q$ limit. Recall that the $ U_{Q}(\hat A(n))$ algebra is
 generated by the set of generators 
 $ \{e_{i}, f_{i}, k_{i}^{\pm} = Q_{i}^{\pm h_{i}} = Q^{d_i ~\pm h_{i}},~~ 0 \le i \le n \}$  satisfying the following relations:
 \begin{equation} \label{e39}
 \begin{array}{c}
 \lbrack e_{i}, f_{j} \rbrack = \delta_{ij} \frac{k_{i} - k_{i}^{-1}}{Q_{i} -Q_{i}^{-1}}   \\
 k_{i}e_{j}k_{i}^{-1}= Q_{i}^{a_{ij}}e_{j},~~k_{i}f_{j}k_{i}^{-1}= Q_{i}^{-a_{ij}}f_{j}\\
 k_{i}k_{i}^{-1}=k_{i}^{-1}k_{i}=1,~~k_{i}k_{j}=k_{j}k_{i}
 \end{array}
 \end{equation}
 and the quantum Serre relations described by the expressions:
 \begin{equation} \label{e40}
 \begin{array}{c}
  \sum_{0\le p \le 1-a_{ij}} (-1)^{p} 
 \left[  \begin{array}{clcr}
 &1-a_{ij}\\ &p \end{array}  \right]_{Q_{i}}e_{i}^{1-a_{ij}-p} e_{j} e_{i}^{p}= 0\\
 \sum_{0\le p \le 1-a_{ij}} (-1)^{p} \left[  \begin{array}{clcr}
 &1-a_{ij} \\&p \end{array}  \right]_{Q_{i}}f_{i}^{1-a_{ij}-p} f_{j} f_{i}^{p} =0
 \end{array}
  \end{equation}
 In equations (\ref{e39}) ,(\ref{e40}) $a_{ij}$ is the ij-element of $n\times n$ generalised Cartan matrix:
 $$
  \hat A_{n} = \left(\begin{array}{rrrrrrrrrrrr}
 2 & -1 & 0 & \cdots & \cdots &&&  -1 \\
 -1 & 2 & -1  &&&&& 0 \\
 0 & -1 & \ddots & \ddots & \ddots &&& \vdots \\
 \vdots & 0 & \ddots & \ddots &   \ddots &&&0\\
 \vdots  &&&&&&&\\
 &&&&&& 2 &-1\\
  -1&0& &&\cdots&0&-1&2\\
 
 \end{array}\right)
 $$\\
 and $(d_i)$ are the non zero integers such that $d_i a_{ij} = a_{ij} d_i$. The quantity $\left[  \begin{array}{clcr}
 &m\\ &n \end{array}  \right]_{Q_{i}}$ in equation (\ref{e39}) is defined by:
 \begin{equation} \label{e41}
 \left[  \begin{array}{clcr}
 &m\\ &n \end{array}  \right]_{Q_{i}}
 = \frac{ \lbrack m \rbrack_{Q_{i}}! }{ \lbrack m -n \rbrack_{Q_{i}}! ~~~\lbrack n \rbrack_{Q_{i}}!}
 \end{equation}
 with
  $$\lbrack H_{i} \rbrack_{Q_{i}}= \frac{Q_{i}^{H_{i}}- Q_{i}^{-H_{i}}}{Q_{i} -Q_{i}^{-1}}$$

 The quantum affine algebra $U_{Q}(\hat A_{n})$ admits two Q-oscillator representations: bosonic and fermionic. in the bosonic realization, the generators of $U_{Q}(\hat A_{n})$ algebra can be constructed, by introducing
 (n +1) $Q$-deformed bosons as follows.
 \begin{equation} \label{e38}
 \begin{array}{c}
 e_{i}= B_{i}^{-} B_{i+1}^{+},~~~ 1 \le i \le n  \\
 f_{i}= B_{i}^{+} B_{i+1}^{-},~~~ 1 \le i \le n  \\
 k_{i}= Q^{-N_{i}+ N_{i+1}},~~~ 1 \le i \le n  \\
 e_{0}= B_{n+1}^{-} B_{1}^{+},~~~ \\
 f_{0}= B_{1}^{-} B_{n+1}^{+},~~~ \\
 k_{0}= Q^{N_{1}- N_{n+1}},~~~ \\
 \end{array}
 \end{equation}
  The fermionic realization is given by:
 \begin{equation} \label{e381}
 \begin{array}{c}
 e_{i}= F_{i}^{+} F_{i+1}^{-},~~~ 1 \le i \le n  \\
 f_{i}= F_{i}^{-} F_{i+1}^{+},~~~1 \le i \le n   \\
 k_{i}= Q^{N_{i}- N_{i+1}},~~~ 1 \le i \le n  \\
 e_{0}= F_{n+1}^{+} F_{1}^{-},~~~ \\
 f_{0}= F_{1}^{+} F_{n+1}^{-},~~~ \\
 k_{0}= Q^{-N_{1}+ N_{n+1}},~~~ \\
 \end{array}
 \end{equation}

 At this stage we investigate the limit $Q \to q$ of the quantum affine algebra $U_{Q}(\hat A_{n}) $. As already mentioned in the introduction, our analysis is based on the Q-oscillator representation. Therefore all  results obtained are specific to the use of Q-Schwinger realization.
 In the $Q \to q$, the splitting of Q-deformed bosons leads to classical bosons $ \{b_{i}^{+}, b_{i}^{-}, N_{b_{i}},~~~(1 \le i \le n) \} $ given by the equation (\ref{e16})and k-fermionic operators $ \{A_{i}^{+}, A_{i}^{-}, N_{A_{i}},~~~~(1 \le i \le n) \}$ defined by equations (\ref{e17}). From the classical bosons, we define the operators
 \begin{equation} 
 \begin{array}{c}\label{e44}
 e_{i}= b_{i}^{-} b_{i+1}^{+},~~~ \\
 f_{i}= b_{i}^{+} b_{i+1}^{-},~~~ \\
 h_{i}= -N_{b_{i}}+ N_{b_{i+1}},~~~ \\
 \end{array}
 \end{equation}
 for $i=1,..., n$ and  
 \begin{equation}\label{e441}
 \begin{array}{c}
 e_{0}= b_{1}^{-} b_{n+1}^{+},~~~ \\
 f_{0}= b_{1}^{+} b_{n+1}^{-},~~~ \\
 h_{0}= N_{b_{1}}- N_{b_{n+1}},~~~ 
 \end{array}
 \end{equation}
 
 The set $\{e_{i}, f_{i},k_{i},~~ 0 \le i \le n \}$ generate the classical algebra $U(\hat A(n))$. From the remaining operators $ \{A_{i}^{+}, A_{i}^{-}, N_{A_{i}},~~~~(1 \le i \le n+1) \}$, one can realize the $U_{q}(\hat A(n))$ algebra. Indeed, the generators defined by
 \begin{equation} 
 \begin{array}{c}
 E_{i}= A_{i}^{-} A_{i+1}^{+},~~~ (1 \le i \le n)\\
 F_{i}= A_{i}^{+} A_{i+1}^{-},~~~ (1 \le i \le n)\\
 K_{i}= q^{-N_{A_{i}}+ N_{A_{i+1}}},~~~ (1 \le i \le n)\\
 E_{0}= A_{1}^{-} A_{n+1}^{+},~~~ \\
 F_{0}= A_{1}^{+} A_{n+1}^{-},~~~ \\
 K_{0}= q^{N_{A_{1}}- N_{A_{i+1}}},~~~ 
 \end{array}
 \end{equation}
 generate the $U_{q}(\hat A(n))$ algebra which is the same version of $U_{Q}(\hat A_{n})$ obtained by simply setting $Q=q$, rather than by taking the limit as above.\\
 Due to the commutativity  of elements  of  $U_{q}(\hat A_{n})$ and $U(\hat A_{n})$,   we obtain the following decomposition of the quantum affine algebra $U_{q}(\hat A_{n})$:
 
 $$ \lim_{Q \to q } ~~~U_{Q}(\hat A_{n}) = U_{q}(\hat A_{n}) \otimes U(\hat A_{n}).$$
 in the bosonic realization. \\
 To end this section, we discuss the equivalence between   $U_{Q}(\hat A(n))$ and $U(\hat A(n))$ algebras in the fermionic construction. Indeed,  We have discussed in the second section how one can identify the conventional fermions with Q-deformed fermions. There have an equivalence between these two objects. Consequently, due to this equivalence, it is possible to construct Q-deformed affine algebras  $U_{Q}(\hat A_{n})$ using ordinary fermions. It is also possible to construct the  affine algebra $\hat A_{n}$  by considering Q-deformed fermions. So, in the fermionic realization we have equivalence between  $U(\hat A_{n})$ and $U_{Q}(\hat A_{n})$. To be more clear, we consider the
 $U_{Q}(\hat A_{n})$  in the Q-fermionic representation. The generators are given by:
 \begin{equation} 
 \begin{array}{c}
 e_{i}= F_{i}^{-} F_{i+1}^{+},~~~ 1 \le i \le n  \\
 f_{i}= F_{i}^{+} F_{i+1}^{-},~~~ 1 \le i \le n  \\
 k_{i}= Q^{N_{F_{i}}- N_{F_{i+1}}},~~~ 1 \le i \le n  \\
 e_{0}= F_{n+1}^{+} F_{1}^{-},~~~ \\
 f_{0}= F_{1}^{+} F_{n+1}^{-},~~~ \\
 k_{0}= Q^{-N_{F_{1}}+ N_{F_{n+1}}},~~~ \\
 \end{array}
 \end{equation}
 due to equivalence fermion - Q-fermion, the operators $f_{i}^{-}, f_{i}^{-}$ are defined as a constant multiple of conventional fermion operators, i.e,
 \begin{equation} 
 \begin{array}{c}
 f_{i}^{-}= Q^{- \frac{N_{F_{i}}}{2}} F_{i}^{-}\\
 f_{i}^{+}=F_{i}^{+} Q^{- \frac{N_{F_{i}}}{2}} 
 \end{array}
 \end{equation}
 from which we can realize the generators:
 \begin{equation} 
 \begin{array}{c}
 E_{i}= f_{i}^{-} f_{i+1}^{+},~~~ 1 \le i \le n \\
 F_{i}= f_{i}^{+} f_{i+1}^{-},~~~ 1 \le i \le n \\
 H_{i}=N_{f_{i}}- N_{f_{i+1}},~~~ 1 \le i \le n \\
 E_{0}= f_{n+1}^{+} f_{1}^{-},~~~ \\
 F_{0}= f_{1}^{+} f_{n+1}^{-},~~~ \\
 H_{0}= - N_{f_{1}}+ N_{f_{n+1}},~~~ \\
 \end{array}
 \end{equation} 
 The set $\{ E_{i}, F_{i}, H_{i} ~~~~,0 \le i \le n \}$ generate the classical affine algebra $U(\hat A_{n})$.\\
 
 \section{ Quantum affine Superalgebra $U_{Q}(\hat A(m,n))$ at Q a root of unity }
 
 Let $ Q \in C-\{ 0 \}$ be the deformation parameter. we shall use also $Q_{i} = Q^{d_{i}}$ with $d_{i}$ are  numbers , that symmetries the Cartan matrix $(a_{ij})$. The quantum affine superalgebra
  $U_{Q}(\hat A(m,n))$ is described in the Serre-Chevalley basis in terms of the simple root  $ e_{i}, f_{i}$ and Cartan generators $h_{i}$ where $ i = 1,....,m+n-1 $ 
 which satisfy the following super-commutations relations
 
 \begin{equation} \label{e444} 
 \begin{array}{c}
 \lbrack e_{i}, f_{j} \rbrack = \delta_{ij} 
 \frac{Q^{d_{i} h_{i}} - Q^{- d_{i} h_{i}}}{Q_{i} -Q_{i}^{-1}}   \\
 \lbrack h_{i}, h_{j} \rbrack =0  \\
 \lbrack h_{i}, e_{j} \rbrack = a_{ij} e_{j},~~ \lbrack h_{i}, f_{j} \rbrack = -a_{ij} f_{j}\\
 \lbrack e_{i}, e_{i} \rbrack = \lbrack f_{i}, f_{i} \rbrack = 0,~~ \\mbox{if} ~~a_{ii} = 0
 \end{array}
 \end{equation}
 with the bracket $\lbrack, \rbrack $ is the $Z_{2}$-graded one
 $$\lbrack X, Y  \rbrack = X Y - (-1)^{deg(X) deg(Y)} Y X. $$
 In the equation (\ref{e444}), $(a_{ij})$ is the element of  the following  Cartan
 matrix:
 $$
  \hat A(m,n) = \left(\begin{array}{rrrrrrrrrrrr}
 0 & 1 & 0 & \cdots & \cdots &&&& \cdots & \cdots & 0 & -1 \\
 -1 & 2 & -1 & 0 &&&&&&&& 0 \\
 0 & -1 & \ddots & \ddots & \ddots &&&&&&& \vdots \\
 \vdots & 0 & \ddots && \ddots & 0 &&&&&& \vdots \\
 \vdots  && \ddots & \ddots & \ddots & -1 & \ddots &&&&& \\
 &&& 0 & -1 & 2 & -1 & \ddots &&&& \\ 
 &&&& \ddots & -1 & 0 & 1 & \ddots &&& \\
 &&&&& \ddots & -1 & 2 & -1 & 0 && \vdots \\
 \vdots &&&&&& \ddots & -1 & \ddots & \ddots & \ddots & \vdots \\
 \vdots &&&&&&& 0 & \ddots && \ddots & 0 \\
 0 &&&&&&&& \ddots & \ddots & \ddots & -1 \\
 -1 & 0 & \cdots & \cdots && \cdots & \cdots & 0 & \cdots & 0 & -1 & 2 \\
 \end{array}\right)
 $$
 
  It is convenient to introduce the quantities $k_{i} = Q_{i}^{d_{i}h_{i}}$ in terms of which the defining relations ( \ref{e444}) become 
 \begin{equation} 
 \begin{array}{c}
 \lbrack e_{i}, f_{j} \rbrack = \delta_{ij} 
 \frac{k_{i} - k_{i}^{-1}}{Q_{i} -Q_{i}^{-1}}   \\
 k_{i}e_{j}k_{i}^{-1}= Q_{i}^{a_{ij}} e_{j},~~k_{i}f_{j}k_{i}^{-1}= Q_{i}^{-a_{ij}} f_{j}\\
 k_{i}k_{i}^{-1}=k_{i}^{-1}k_{i}=1,~~k_{i}k_{j} = k_{j}k_{i}\\
 \lbrack e_{i}, e_{i} \rbrack = \lbrack f_{i}, f_{i} \rbrack = 0, \mbox{if} ~~~a_{ii} = 0
 \end{array}
 \end{equation}
 Further the quantum affine superalgebra
  $U_{Q}(\hat A(m,n))$ generators obey to generalised Serre relations .The latter's  are most simply presented in terms of the following rescaled generators \cite{r22}:
 $$ \xi_{i} = e_{i} k_{i}^{-\frac{1}{2}},~~\zeta_{i} = f_{i} k_{i}^{-\frac{1}{2}} $$
 they then take the form
 \begin{equation} 
 \begin{array}{c}
 (ad_{Q} \xi_{i})^{1- \tilde a_{ij}} \xi_{j} = 0,~~i \ne j, \\
 (ad_{Q} \zeta_{i})^{1- \tilde a_{ij}} \zeta_{j} = 0,~~i \ne j.
 \end{array}
 \end{equation}
 
 where $\tilde a_{ij}$ matrix is obtained from the non-symmetric Cartan matrix $ a_{ij}$  by substituting
 -1 for the strictly positive elements in the rows with $0$ on the diagonal entry.
 The quantum adjoint action $ad_{Q}$ can explicitly  written in terms of the coproduct and the antipode as :
 $$ (ad_{Q}X)Y = (-1)^{deg(X_{(2)}). deg(Y) } X_{(1)} Y S(X_{(2)}) $$
 with $ \Delta(X) = X_{(1)} \otimes X_{(2)}$.
 and some supplementary relations for $i$ such that $a_{ii} = 0$.
 $$ \lbrack  [ e_{i -1}, e_{i} ]_{Q} , [ e_{i }, e_{i +1} ]_{Q} \rbrack  = 0$$
 $$ \lbrack [ f_{i -1}, f_{i} ]_{Q} , [ f_{i }, f_{i +1} ]_{Q}  \rbrack  = 0 $$
 
 The universal quantum affine Lie superalgebra $U_{Q}(\hat A(m,n))$ is endowed with a Hopf superalgebra structure with coproduct: 
 $$ \Delta(k_{i}) = k_{i} \otimes  k_{i}$$
 $$ \Delta(e_{i}) = e_{i} \otimes k_{i}^{\frac{1}{2}} + k_{i}^{-\frac{1}{2}} \otimes e_{i} $$
 $$ \Delta(f_{i}) = f_{i} \otimes k_{i}^{\frac{1}{2}} + k_{i}^{-\frac{1}{2}} \otimes f_{i} $$
  counit:  
 $$ \epsilon (k_{i}) = 1,~~\epsilon(e_{i}) = \epsilon(f_{i}) = 0 $$
  and antipode :
 $$ S(k_{i}) = -k_{i},~~~~ S(e_{i}) =  - Q_{i}^{\frac{a_{ii}}{2}} e_{i},~~~S(f_{i}) =- Q_{i}^{\frac{- a_{ii}}{2}} f_{i}$$
 We shall give now the $Q$-oscillator representation of the quantum affine superalgebra 
 $U_{Q}(A(m,n))$. We shall provide explicit expressions for corresponding generators as linear and bilinear in Q-deformed bosonic and fermionic oscillators.
 The quantum affine superalgera $U_{Q}(\hat A(m,n))$ can be realized simply by (m+1)Q-deformed fermions and (n+1) Q-bosons. Explicitly the generators of $\hat A_{Q}(m,n)$ are given by:
 \begin{equation} 
 \begin{array}{c}
 e_{i}= F_{i}^{+} F_{i+1}^{-},~~~1 \le i \le m \\
 f_{i}= F_{i}^{-} F_{i+1}^{+},~~~1 \le i \le m \\
 k_{i}= Q^{  (N_{F_{i}}- M_{F_{i+1}})},~~~ 1 \le i \le m\\
 e_{m+1}= F_{m+1}^{+} B_{1}^{-},~~f_{m+1}= F_{m+1}^{+} B_{1}^{+},~~k_{m+1}=  Q^{  (N_{F_{m+1}}+ N_{B_{1}}) }\\
 e_{m+j}= B_{j-1}^{+} B_{j}^{-},~~~2 \le j \le n+1 \\
 f_{m+j}= B_{j-1}^{-} B_{j}^{+},~~~2 \le j \le n+1 \\
 k_{m+j}= Q^{  (N_{B_{j-1}}- N_{B_{j}})},~~~ 2 \le j \le n+1 \\
 
 e_{0}= B_{n+1}^{+} F_{1}^{-}, \\
 f_{0}= F_{1}^{+} B_{n+1}^{+}, \\
 k_{0}= Q^{  (N_{F_{n+1}}+ N_{B_{1}})}
 \end{array}
 \end{equation} 
 Due to the property of Q-boson decomposition in the $Q \to q$ limit, each Q-boson $\{B_{i}^{-}, B_{i}^{+}, N_{B_{i}} \}$ reproduce an ordinary boson $\{b_{i}^{-}, b_{i}^{+}, N_{b_{i}} \}$ and a k-fermion operator $\{A_{i}^{-}, A_{i}^{+}, N_{A_{i}} \}$. In the limit the Q-fermions become q-fermions which are objects equivalents to conventional fermions $\{f_{i}^{-}, f_{i}^{+}, N_{f_{i}} \}$. From the classical bosons $\{b_{i}^{-}, b_{i}^{+}, N_{b_{i}} \}$ and conventional fermions $\{f_{i}^{-}, f_{i}^{+}, N_{f_{i}} \}$, one can realize 
  the classical affine algebra $U(\hat A(m,n))$:
 \begin{equation} 
 \begin{array}{c}
 E_{i}= f_{i}^{+} f_{i+1}^{-},~~~1 \le i \le m \\
 F_{i}= f_{i}^{-} f_{i+1}^{+},~~~1 \le i \le m \\
 H_{i}=  N_{f_{i}}- N_{f_{i+1}} ,~~~ 1 \le i \le m\\
 E_{m+1}= f_{m+1}^{+} b_{1}^{-},~~F_{m+1}= f_{m+1}^{+} b_{1}^{+},~~H_{m+1}=   N_{f_{m+1}}+ N_{b_{1}} \\
 E_{m+j}= b_{j-1}^{+} b_{j}^{-},~~~2 \le j \le n \\
 F_{m+j}= b_{j-1}^{-} b_{j}^{+},~~~2 \le j \le n \\
 H_{m+j}= N_{b_{j-1}}- N_{b_{j}},~~~ 2 \le j \le n \\
 E_{0}= b_{n+1}^{+} f_{1}^{-}, \\
 F_{0}=f_{1}^{+} b_{n+1}^{+}, \\
 H_{0}=N_{f_{n+1}}+ N_{b_{1}}
 \end{array}
 \end{equation} 
  From the operators $\{A_{i}^{-}, A_{i}^{-}, N_{A_{i}} \}$ we construct the generators 
 \begin{equation} 
 \begin{array}{c}
 e_{i}= A_{i}^{-} A_{i+1}^{+},~~~ 1 \le i \le n+1\\
 f_{i}= A_{i}^{+} A_{i+1}^{-},~~~ 1 \le i \le n+1\\
 k_{i}= q^{-N_{A_{i}}+ N_{A_{i+1}}},~~~ 1\le i \le n+1\\
 e_{0}= A_{n+1}^{-} A_{1}^{+}, \\
 f_{0}=  A_{n+1}^{+} A_{1}^{-}, \\
 k_{0}= q^{  N_{A_{n+1}}+ N_{A_{1}}}
 \end{array}
 \end{equation}
 for $1 \le i \le n+1$, which generates the algebra $U_{q}(\hat A_{n})$. It is easy to verify that $U_{q}(\hat A(n))$ and $\hat A(m,n)$ are mutually commutative. As results, we have the following decomposition of quantum superalgebra $A_{Q}(m,n)$ in the $Q \to q$ limit 
 $$  lim_{Q \to q} U_{Q}(\hat A(m,n)) = U(\hat A(m,n)) \otimes U_{q}(\hat A(n)).  $$
 
 \section{Conclusion}
 
  We have presented the general method leading to the investigation the $Q \to q=  e^{\frac{\rm 2\pi i}{\rm  k}}$ limit 
 of the  quantum affine algebra $U_{Q}(\hat A_{n})$  and quantum affine superalgebra  $U_{Q}(\hat A(m,n))$ based on the decomposition of $Q$-bosons in this limit.
 We note that $Q$-oscillator realization is crucial in this manner of spitting in this paper.  We believe that the techniques and formulae, used here,
 will be useful foundation to extend this study to all quantum affine algebras, quantum affine superalgebras and $Q$-deformed exceptional  Lie algebras
 and superalgebras. 
 \pagebreak
 
 \end{document}